\documentclass[3p,twocolumn]{elsarticle}

\usepackage{amssymb}

\usepackage[figuresright]{rotating}

\newcommand{\Pom}{I$\!$P}            

\begin{document}

\begin{frontmatter}




\title{Minimum Bias, MPI and DPS, Diffractive and Exclusive measurements at CMS}


\author{Dipanwita Dutta on behalf of CMS Collaboration}

\address{Nuclear Physics Division, Bhabha Atomic Research Centre, Mumbai-400085, India.}

\begin{abstract}
We present recent results on Minimum Bias, MPI and DPS,  Diffractive and Exclusive studies
using data collected during Run 1 of the LHC. The measurements include data collected in 
pp collisions at $\sqrt s =$ $7$, and $8$ TeV by the CMS Collaboration.
Double parton scattering is investigated in several final states including
vector bosons and jets, and the effective cross section results are compared to
other experiments and to MPI models tuned to recent underlying event measurements at CMS. 
Inclusive diffractive cross sections are discussed and compared to models, while searches
and measurements of central exclusive processes are presented.
The results from the first combined measurement by the CMS+TOTEM collaborations
of the pseudorapidity distribution of charged particles at 8 TeV are also discussed, 
and are compared to models and to lower energy measurements.

\end{abstract}


\end{frontmatter}


\section{Introduction}
Forward, diffractive and exclusive physics cover a wide range of
subjects, including low-x QCD, underlying event and
multiple interactions characteristics,  and 
central exclusive process.
With  excellent performance the Compact Muon Solenoid (CMS) experiment \cite{ref0}
has made a number of significant observations in diffractive
and exclusive processes and hence to probe the Standard model
in a unique way.
The particle production in pp collisions at LHC,
will allow to test the fundamental aspect of QCD,
namely the interplay between soft and hard contributions to an interaction. 
Its good understanding is crucial for the proper modeling
 of the final state of Minimum-Bias events,
and can help improve the simulation of e.g. the underlying event, pile-up events, and the measurement
of the machine luminosity at the LHC.
In this paper, we present the recent CMS results on diffraction, 
forward physics and soft QCD, and discuss their comparison
to predictions of various theoretical models.

\begin{figure}
\includegraphics[width=72mm]{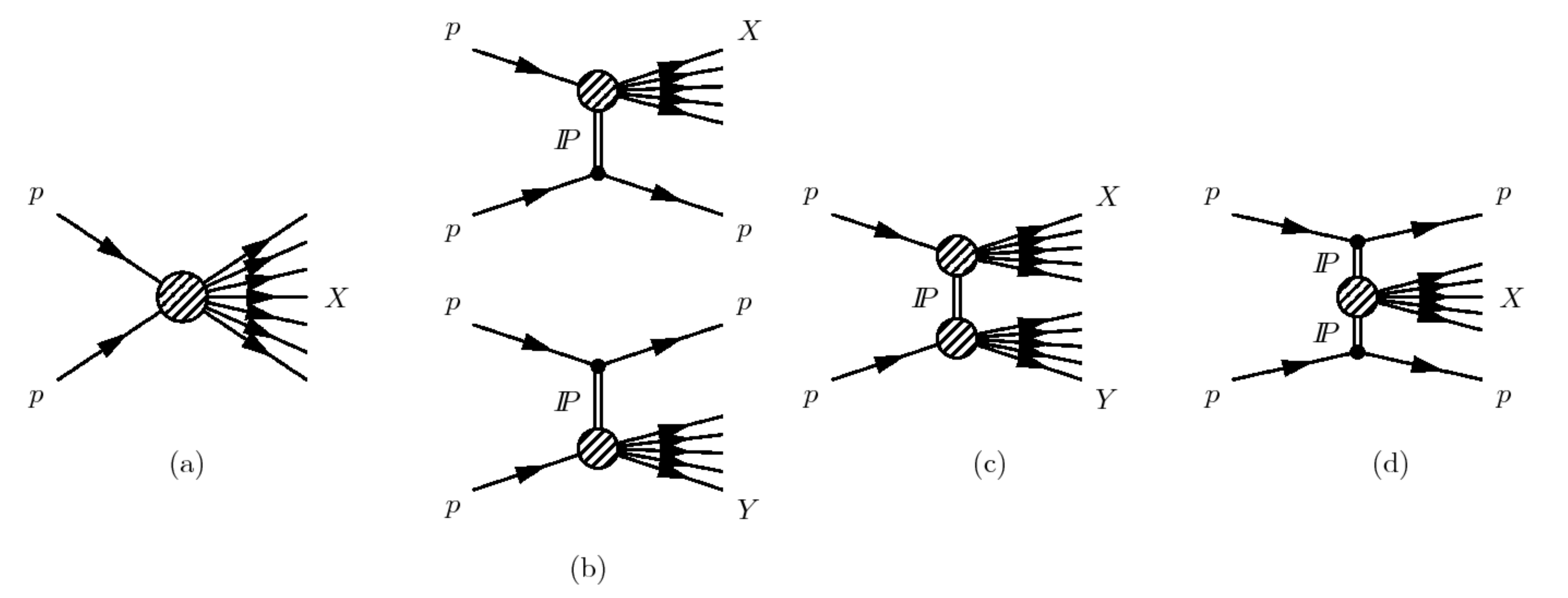}
\vspace{-0.4cm}
\caption{
Schematic diagrams of (a) non-diffractive, $pp \rightarrow X$, and diffractive processes with
(b) single dissociation,
$pp \rightarrow X p$ or $pp \rightarrow pY$, (c) double-dissociation, $pp \rightarrow XY$,
and (d) central dissociation,
$pp \rightarrow pX p$. The X(Y) represents a dissociated-proton
or a centrally-produced hadronic system.}
\label{diff}
\end{figure}
\section{Diffractive processes}
Diffractive interactions are characterized by the presence of at least one non-exponentially
suppressed large rapidity gap (LRG) in the final state. LRG is defined as a region in pseudorapidity
devoid of particles is presumed to be formed by a  color-singlet exchange with vacuum
quantum numbers, referred to as Pomeron (\Pom) exchange.
Inclusive (soft) diffractive interactions (with no hard scale) cannot be calculated within perturbative
QCD (pQCD), and traditionally have been described by models based on Regge theory. Model
predictions generally differ when extrapolated from pre-LHC energies (e.g. 1.96 TeV) to 7 TeV
at LHC. Thus experimental results at LHC provide important
input for tuning various models and current event generators.
Fig.~\ref{diff} shows the main types of diffractive
processes: single dissociation (SD), double dissociation (DD) and central dissociation (CD).

\begin{figure}
\includegraphics[width=75mm]{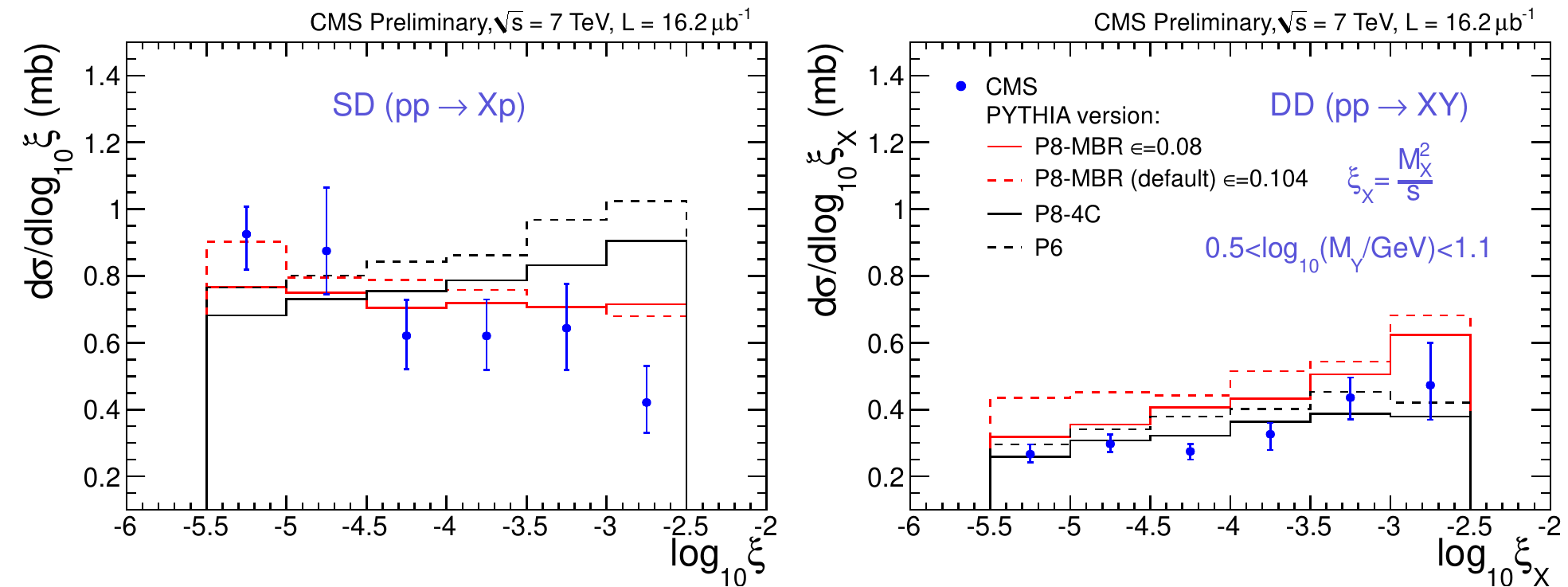}
\vspace{-0.4cm}
\caption{
The SD (left) and DD (right) cross sections as a function of $\xi$ compared to PYTHIA6,
 PYTHIA8-4C and PYTHIA8-MBR MC predictions.}
\label{xi_dist}
\end{figure}

Diffractive cross sections have been measured with CMS \cite{ref1}
using the low-pileup 2010 data of pp collision at $\sqrt s = 7$ TeV,
corresponding to an integrated luminosity of $16.2$ $\mu b^{-1}$.
The SD and DD events 
were separated using the CASTOR calorimeter,
which covers the very forward region of
the experiment, $-6.6 < \eta < -5.2$.
Minimum bias events were selected by requiring a signal
above noise level in any of the BSC (Beam
Scintillator Counter) devices ($3.2 < | \eta | < 4.7$)
and the presence of at least two energy deposits
in the central CMS detector ($| \eta | < 4.7$).
Diffractive events
were selected by requiring the presence of a forward rapidity gap reconstructed at the edge
of the central detector or central gap. The forward gap on the positive (negative) side was
reconstructed in terms of the variable $\eta_{max}$ ($\eta_{min}$) 
defined as the highest (lowest) $\eta$ of a particle
reconstructed in the detector. The central gap was reconstructed as 
$\Delta \eta^{0} = \eta^{0}_{max} - \eta^{0}_{min}$,
 with $\eta^{0}_{max}$ ($\eta^{0}_{min}$) defined as the 
closest-to-zero $\eta$ of a particle reconstructed on the positive
(negative) $\eta$-side of the central detector, 
with an additional requirement of activity on both
sides of the detector.
The event sample after the $\eta_{min} > -1$ selection 
was used to extract SD and
DD cross sections. Subsamples enhanced in SD
and DD events were selected by requiring an absence or
a presence of an energy deposit in the CASTOR
calorimeter. The differential SD cross section as a function of  $\xi$
(an incoming-proton momentum loss), and
the differential DD cross section as a function of 
$\xi_{X} = M^{2}_{X}/s$ for $0.5 < \log_{10}(M_{Y}/GeV) < 1.1$
(CASTOR acceptance), after subtracting the background contribution
to the signal (DD to SD and ND to DD), are
shown in Fig.~\ref{xi_dist} (left) and (right), 
respectively. Results are compared to MC models.

The diffractive-event generation in MBR is based on
a phenomenological renormalized Regge theory model \cite{ref1a}.
The predictions of PYTHIA8-MBR are shown for two values of the 
parameter of the Pomeron trajectory $\alpha (t) = 1 + \epsilon + \alpha^{\prime}(t)$,
$\epsilon= 0.08$ and $\epsilon= 0.104$.
Both values describe the measured SD cross section within uncertainties, while
the DD data favor the smaller value of $\epsilon= 0.08$.
The predictions of PYTHIA8-4C and PYTHIA6 describe well the
measured DD cross section, but fail to describe the falling behavior of the data. 
The total measured SD cross section 
integrated over the region $−5.5 < \log_{10} \xi < −2.5$ ($12 \leq M_{X} \leq 394$ GeV)
is $\sigma_{vis}^{SD}=4.27 \pm 0.04(stat.)^{+0.65}_{-0.58}(syst.)$ mb.

CMS with a close cooperation with TOTEM experiment, located at the
same interaction point provides almost full
coverage in pseudorapidities for charged and neutral particles.
The CMS and TOTEM collaborations have measured \cite{ref1b} 
pseudorapidity distributions of charged particles, $dN_{ch} /d\eta$, using
the low-pileup 2012 data at $\sqrt s = 8$ TeV
($L = 45 \mu b^{-1}$), recorded during the common CMS and TOTEM
runs with a non standard ($\beta^{*} =90$m) LHC optics configuration.
This is the first result of the combined CMS and TOTEM analysis,
covering the ranges of $|\eta| < 2.2$ and $5.3 < |\eta| < 6.4$,
 respectively. 
\begin{figure}
\begin{center}
\includegraphics[width=55mm]{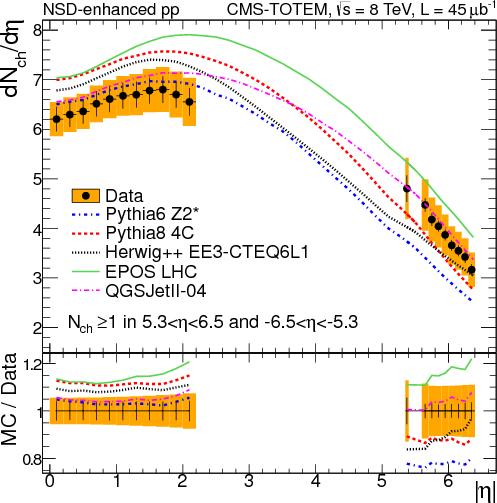}
\vspace{-0.4cm}
\caption{
Charged-particle pseudorapidity distributions from a
NSD-enhanced sample.
}
\label{pseudo_dist}
\end{center}
\end{figure}
Depending on the configuration of the T2 detectors with a signal, events were
categorized into three different samples: (i) an inclusive sample, 
sensitive to $91-96\%$ of the total inelastic
proton-proton cross section, (ii) a sample enhanced in non-single diffractive 
(NSD-enhanced) events, and (iii) a sample enhanced in 
single-diffractive (SD-enhanced) events. 
The measured $dN_{ch} /d\eta$ distributions for 
NSD-enhanced  samples are presented in 
Fig.~\ref{pseudo_dist} showing
that the charged particle density decreases with $\eta$.
The results are compared to the predictions of various
Monte Carlo models: PYTHIA6-Z2*,
PYTHIA8-4C, HERWIG++, EPOS, and QGSJET-II-04. None of the
models provides a consistent description of the measured distributions.

\begin{figure}
\begin{center}
\includegraphics[width=50mm]{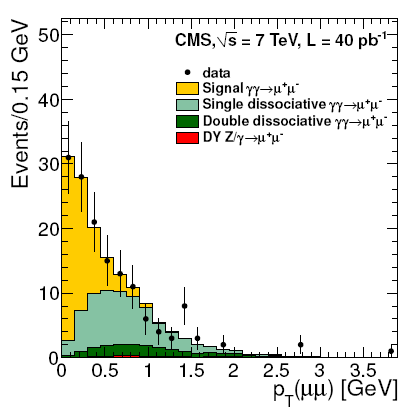}
\includegraphics[width=50mm]{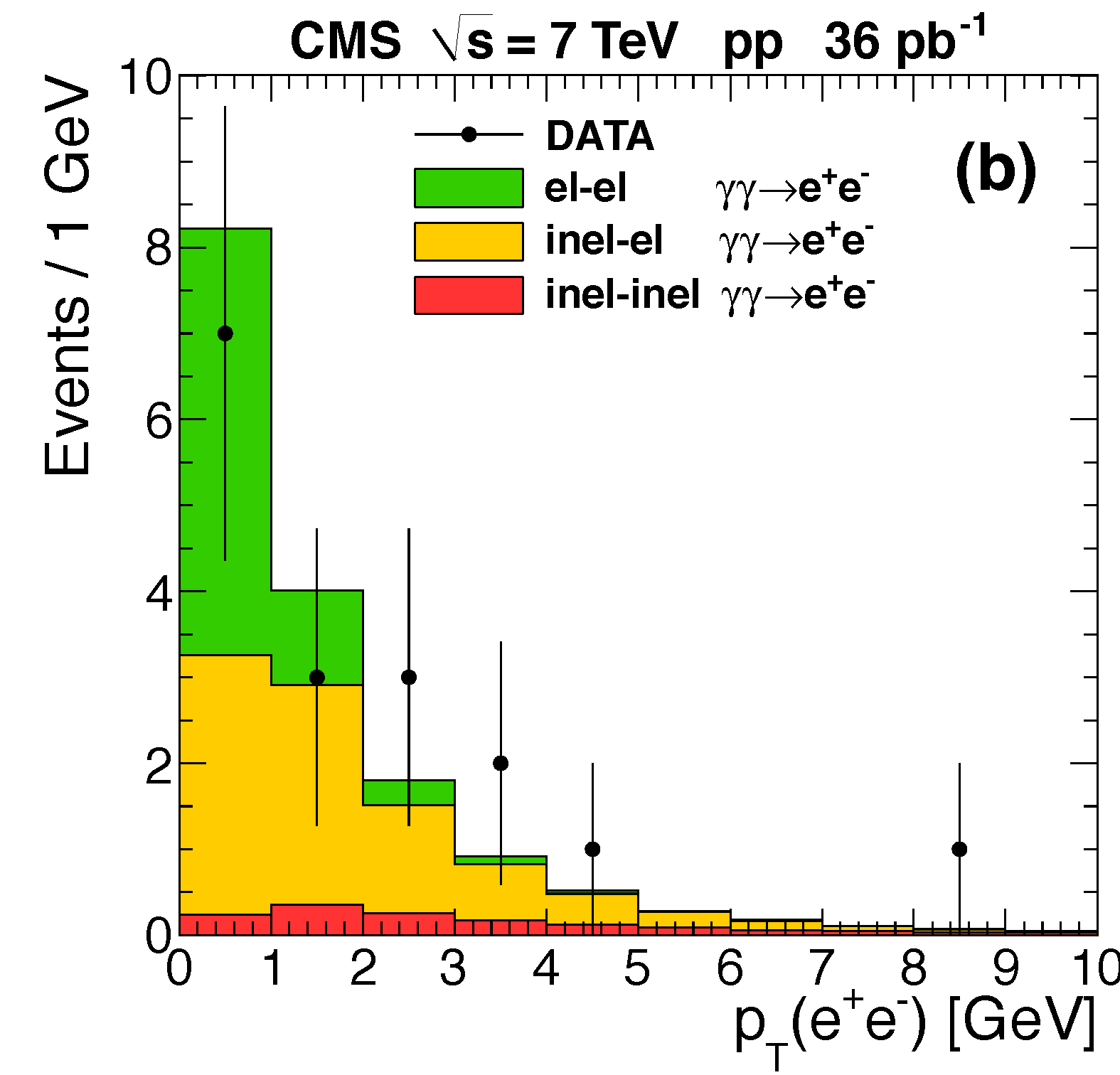}
\vspace{-0.4cm}
\caption{
Muons (top) and electron (bottom) pairs transverse momentum distributions
for the candidates selected in the two-photon production of leptons pairs.
}
\label{gg_mumu}
\end{center}
\end{figure}
\section{Central Exclusive processes}
Another class of processes with a LRG in the final state is central exclusive  process (CEP).
The CEP is a process of the type: $pp \rightarrow p + X + p$ with X being a well defined system e.g. di-lepton
or di-jet.  Exclusive means no additional activity between the outgoing protons and X, thus,
the final state consists of the scattered protons which survive the interaction intact, and of the system
X or its decay products.
In the CEP, 
three distinct processes may be involved, namely photon-photon,
photon-pomeron and pomeron-pomeron interactions. 
The system $X$ is reconstructed in the central
CMS barrel, while forward detectors are used to veto non-exclusive events.

Exclusive di-lepton production
$\gamma \gamma\rightarrow l^{+}l^{-}$ is a nearly pure QED process. Therefore its cross section
 is precisely known. Its measurement at the LHC is an independent cross check of the the
absolute luminosity calibration \cite{ref2,ref3}.
Two different di-lepton exclusive analyses have been performed using the data
collected in 2010 at 7 TeV, namely for the measurements of 
$\gamma \gamma \rightarrow e^{+}e^{-}$~\cite{ref2} 
and of $\gamma \gamma \rightarrow \mu^{+} \mu^{-}$~\cite{ref3}. 
The event selection is requiring two leptons, which are energy or momentum
balanced and back-to-back in the transverse plane. This corresponds to a
$|p_{T} (l^{+}) - p_{T} (l^{-})| < 1$ GeV as well as an acoplanarity
describing the difference in azimuthal angles,
$|1 - \Delta \phi( l^{+}, l^{-} )/\pi| < 0.1$.The dimuon analysis
requires each of the two muons to carry a transverse momentum larger than
$4$ GeV in the range $|\eta(\mu)| < 2.1$. In order to reject the exclusive photoproduction
of the low-mass resonances, an
invariant mass cut is applied $M_{\mu^{+}\mu^{-}} > 11.5$ GeV.
For the dielectron analysis, a electron-positron pair with a transverse energy deposit in the calorimeters
$E_{T} > 5.5$ GeV are selected in the range $|\eta(e)| < 2.5$. 
The selected sample consists of events of exclusive
as well as semi-exclusive dilepton production, in which
the dissociated proton escapes detection in the central detector. 
Fig.~\ref{gg_mumu} shows the distributions
of the $p_T$ of the di-muon pair (top) and di-electrton pair (bottom),
and  compared to LPAIR MC predictions for exclusive
and semi-exclusive production. Good
agreement between data and the simulation is observed.
These two results allow to improve the understanding of this purely electromagnetic
process, by observing $17$ candidates for the dielectron channel and
by a measurement of a production cross-section at $\sqrt s=7$ TeV 
for the dimuon channel:
\begin{eqnarray}
\sigma_(pp \rightarrow p\mu^{+} \mu^{-} p)  = 3.38^{+0.58}_{-0.55} (\mbox{stat.})\nonumber\\
\pm 0.16 (\mbox{syst.}) \pm 0.14 (\mbox{lumi.})~\mbox{pb}. \nonumber
\end{eqnarray}

\begin{figure}
\begin{center}
\includegraphics[width=55mm]{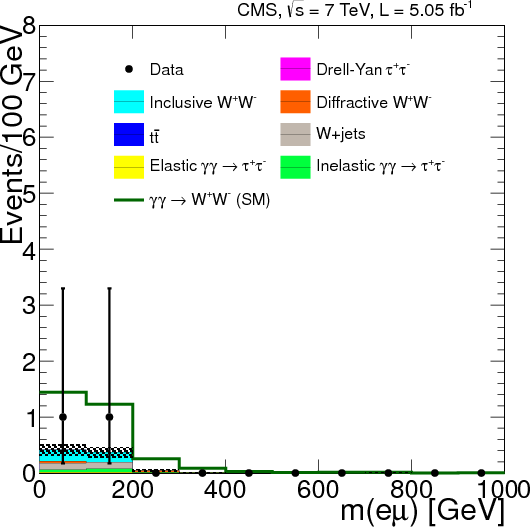}
\vspace{-0.4cm}
\caption{
$\mu$e invariant mass  for events in the signal region with 0 extra tracks on the
 $\mu$e vertex and $p_T(\mu e) > 30$ GeV. The backgrounds (solid histograms) are stacked 
with statistical uncertainties indicated by the shaded region, the signal histogram
(open histogram) is stacked on top of the backgrounds.
}
\label{wwpair}
\end{center}
\end{figure}

Several processes beyond the Standard model predict an anomalous
quartic gauge coupling (AQGC)
which can be translated into a higher production rate, or discrepancies in the kinematic
distributions of multiple final state particles.
A search for exclusive or quasi-exclusive $W^{+} W^{-}$ production by photon-photon
interactions, $pp \rightarrow p^{(\star)} W^{+} W^{-} p^{(\star)}$, at $\sqrt s = 7$ TeV
 is reported using data collected by the
CMS detector with an integrated luminosity of $5.05$ fb$^{-1}$.
 Events are selected by requiring
a $\mu^{\pm} e^{\mp}$ vertex with no additional associated charged tracks 
and dilepton transverse momentum
$p_{T}(\mu^{\pm}e^{\mp})>30$ GeV. 
Two events passing all selection requirements are observed
in the data, compared to a standard model expectation of $2.2\pm 0.4$
signal events with $0.84 \pm 0.15$ background (Fig.~\ref{wwpair}). The tail of 
the dilepton $p_T$ distribution is studied for deviations
from the standard model. No events are observed with $p_T > 100$ GeV.
Model independent upper limits are computed and compared to predictions
involving anomalous quartic gauge couplings \cite{ref4}.

\section{Underlying events, MPI and DPS}
In a proton-proton scattering the hadronic final state
can be described as the superposition of different contributions.
Most of the inelastic particle production 
can be described in a picture where an event is a combination of
hadronic jets, originating from hard parton-parton
interactions with exchanged momenta above
several GeV/c and of an underlying event, which, consists
of softer parton-parton interactions and of proton remnants. 
The underlying event (UE) is commonly defined as the set of all
final-state particles that are not associated with the initial
hard-parton scattering. This component is presumably dominated
by perturbative (mini)jets with relatively small transverse
momenta of a few GeV/c, produced in softer multi-parton interactions (MPI) as well as
by soft hadronic strings from the high-rapidity remnants.
\begin{figure}
\begin{center}
\includegraphics[width=65mm]{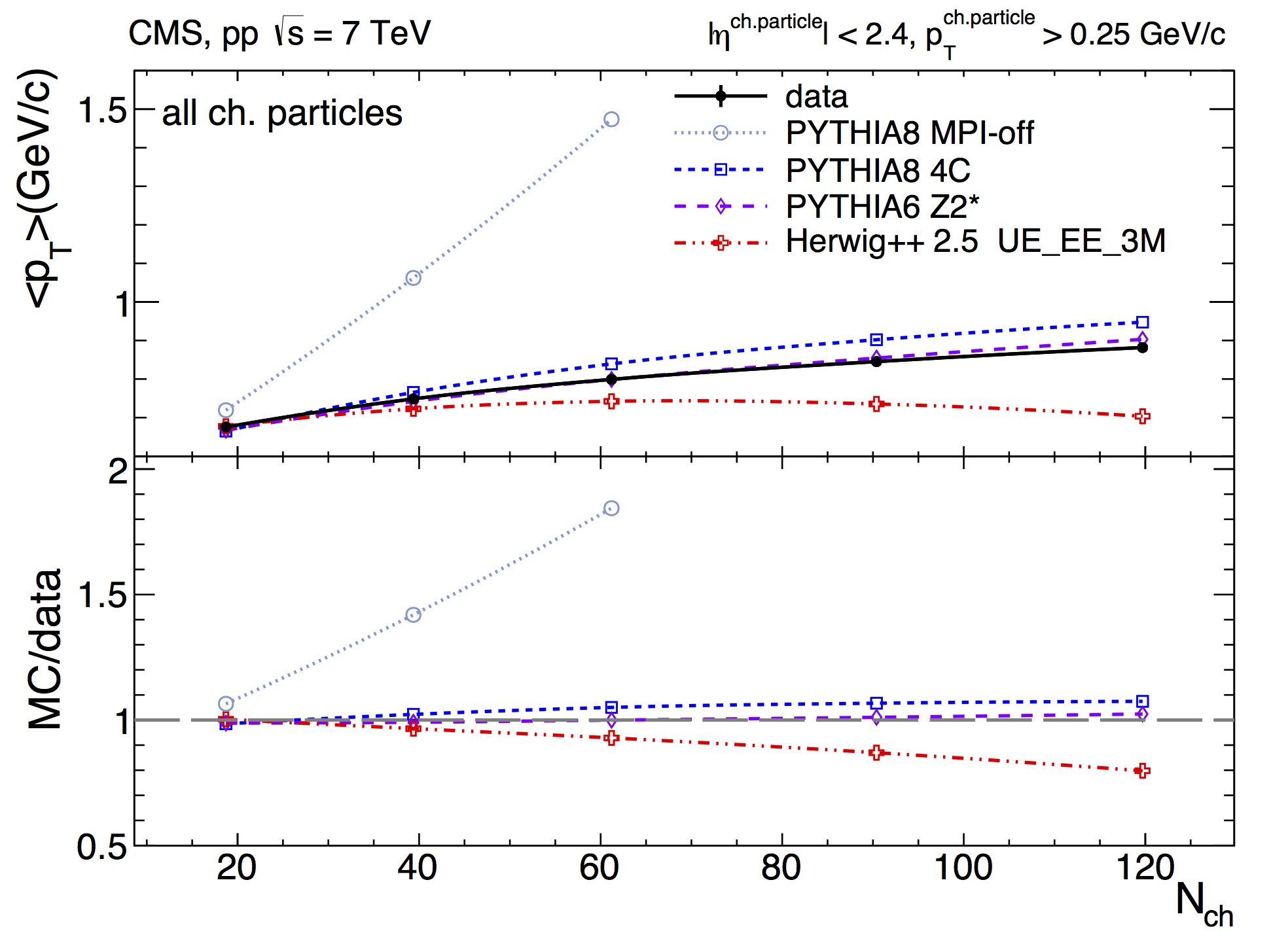}
\vspace{-0.4cm}
\caption{
Mean transverse momentum of inclusive charged tracks with $p_T > 0.25$ GeV/c 
versus the corrected charged-particle multiplicity ($N_{ch}$ within $|\eta|<2.4$). 
} 
\label{mpi}
\end{center}
\end{figure}

Multi-particle production in proton-proton collisions at $\sqrt s = 7$ TeV
are studied as a function of the charged-particle multiplicity, $N_{ch}$ \cite{ref5}.
The produced particles are separated into two classes: those belonging to jets and those belonging to
the underlying event. Charged particles are measured with pseudorapidity $|\eta | < 2.4$
and transverse momentum $p_{T} > 0.25$ GeV/c. Jets are reconstructed from charged particles 
only and having $p_T > 5$ GeV/c. The distributions of jet $p_T$, average
$p_T$ of charged particles belonging to the underlying event or to jets, jet rates, and
jet shapes are presented as functions of $N_{ch}$ and compared to the predictions of the
PYTHIA and HERWIG event generators. 
Fig.~\ref{mpi} shows the distribution of mean transverse momentum
of inclusive charged tracks versus the corrected pp charged-particle multiplicity.
Current event generators tuned to reproduce
the inelastic LHC data cannot describe within a single approach the dependence of various
quantities on event multiplicity. 
For increasing $N_{ch}$, PYTHIA systematically predicts higher jet rates and harder $p_T$
spectra than seen in the data, whereas HERWIG shows the opposite trends.
Predictions of PYTHIA without multi-parton interactions
fail completely to describe the $N_{ch}$ dependence observed in the data,
which demonstrate that  MPI mechanism is critical for reproducing the measured properties of
the jets and UE for moderate and large charged-particle multiplicities.
At the highest multiplicity, the
data-model agreement is worse for most observables, 
indicating the need for further tuning and/or new model ingredients.

LHC probes small values of the momentum fraction $x$ carried by the colliding partons
and the large densities at small-x values increase the probability of having two simultaneous
parton-parton scatterings producing two independently identifiable hard scatterings
in a single interaction. 
A study of double parton scattering (DPS) has been performed \cite{ref6} with W+2-jet events,
using 5 fb$^{-1}$ of the data at $\sqrt s = 7$ TeV. DPS with a W+2-jet final
state occurs when one hard interaction produces a W boson and another
produces a dijet in the same pp collision. Events with a W
boson, reconstructed from the muon of $p_T > 35$ GeV and the missing
transverse energy of $\not{E}_T > 30$ GeV, were required to have exactly
two jets with $p_T > 20$ GeV and $|\eta | < 2$.
Several observables, which are sensitive
to discriminate DPS events from the Single Parton Scattering (SPS) ones,
$\Delta \phi$, $\Delta^{rel}$ $p_T$ and $\Delta S$ (as defined in \cite{ref6}),
have been studied. Corrected distributions are compared with particle 
level predictions of MADGRAPH  MC sample.
\begin{figure}
\begin{center}
\includegraphics[width=55mm]{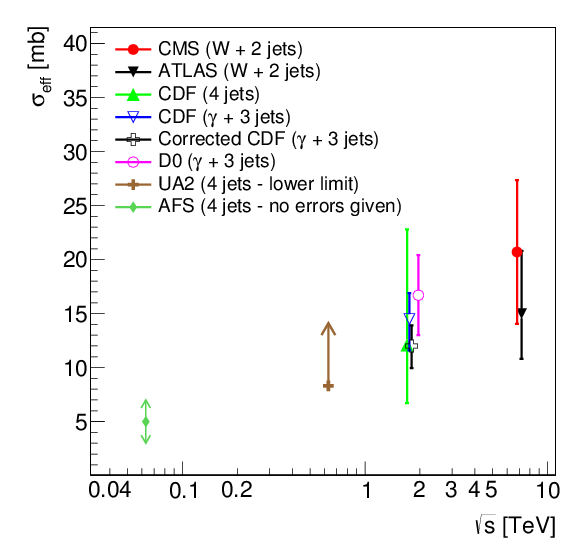}
\vspace{-0.4cm}
\caption{
Center of mass energy dependence of $\sigma_{eff}$ measured by
different experiments using different processes. 
} 
\label{dps}
\end{center}
\end{figure}
The effective DPS cross
section, $\sigma_{eff}$, has been measured using a relation
 $\sigma_{eff} = R . \sigma_{2j} /f_{\mbox{DPS}}$,
where, $R = N_{W+0j}/N_{W+2j}$, and $\sigma_{2j}$ are the ratio of
 W+0-jet to W+2-jet events and the di-jet production cross section, 
 respectively; $f_{\mbox {DPS}}$ is the fraction of DPS events in the 
W+2-jet sample.
The fraction of DPS in W + 2-jet events is extracted with a DPS + SPS template
fit to the distribution of the $\Delta^{rel}$$p_T$ and $\Delta$S observables.
 The obtained value of the DPS fraction is
\begin{eqnarray}
f_{DPS} = 0.055 \pm 0.002 (\mbox{stat.}) \pm 0.014 (\mbox{syst.})\nonumber
\end{eqnarray}
and the effective cross section, 
characterizing the effective transverse area of hard partonic
interactions in collisions between protons, is calculated to be
\begin{eqnarray}
\sigma_{eff} = 20.7 \pm 0.8 (\mbox{stat.}) \pm 6.6 (\mbox{syst.}) \mbox{mb}. \nonumber
\end{eqnarray}
The measured value of the effective cross section is
consistent with the Tevatron and ATLAS results (Fig.~\ref{dps}) .

\section{Summary and outlook}
In this paper, several achievements are being
presented in the experimental search for 
diffractive, exclusive and underlying events at the LHC.
Excellent experimental  measurement of separated SD and DD 
cross-sections are being reported and
compared with revised  PYTHIA version
based on a renormalised Regge theory model.
The results from the first combined measurement by the CMS+TOTEM collaborations
of the pseudorapidity distribution of charged particles at 8 TeV are discussed, 
and are compared to models and to lower energy measurements.
In CEP process, with two candidates on 
$\gamma \gamma \rightarrow W^{+} W^{-} $ process, the 
best limits on the anomalous quartic couplings is 
extracted. The  effective cross section for DPS,
measured with CMS, are also found to be 
consistent with other experimental results.

The results presented in this paper provide the evidence for the
excellent performance of CMS experiment and its potential for future
measurements of diffractive, exclusive and soft QCD physics.




.


\end{document}